\documentstyle[12pt,psfig]{article}
\newcommand{\ba}{\begin{eqnarray}}
\newcommand{\ea}{\end{eqnarray}}
\newcommand{\half}{{\textstyle{\frac{1}{2}}}}
\newcommand{\partialslash}{\partial\hspace{-.5em}/\hspace{.15em}}
\newcommand{\Pslash}{P\hspace{-.5em}/\hspace{.15em}}

\newcommand{\kslash}{k\hspace{-.5em}/\hspace{.15em}}

\newsavebox{\s}
\newsavebox{\tmpbox}
\newcommand{\Slash}[1]{\sbox{\s}{#1} \hbox to \wd\s {#1\hss\hbox to \wd\s{\hss/\hss}}}

\newcommand{\be}{\begin{equation}}
\newcommand{\ee}{\end{equation}}

\newcommand{\kstar}{K_{0}^{*}}
\newcommand{\diag}{{\rm diag}}

\newcommand{\Tr}{{\rm Tr}}
\newcommand{\mev}{{\kern2mm\rm MeV}}
\newcommand{\gev}{{\kern2mm\rm GeV}}

\newcommand{\journal}[5]{#1, {\it #2} {\bf #3}, #5 (#4)}

\begin{document}
\makeatletter
\makeatother

\begin{titlepage}
\title{
\hfill{\normalsize JINR Preprint E2-99-89}\\[2cm]
\bf Excited scalar mesons in a chiral quark model}
\author{M.K.Volkov, V.L.Yudichev}
\date{}
\end{titlepage}

\maketitle
\begin{abstract}
First radial excitations of the
isoscalar and isovector scalar mesons $f_0(400-1200), f_0(980)$
and $a_0(980)$  are investigated in the
framework of a nonlocal version of a chiral quark model of the
Nambu--Jona-Lasinio (NJL) type. It is shown that 
$f_0(1370)$, $f_J(1710)$  and $a_0(1450)$
are the first radially excited states of
$f_0(400-1200)$, $f_0(980)$ and $a_0(980)$ which 
are ground states of the scalar meson
nonet. The mesons' masses and strong
decay widths are calculated. The scalar resonance $f_0(1500)$ is 
supposed to be a glueball. The status of $\kstar(1430)$ is discussed.
\end{abstract}
\vspace{1cm}

\sbox{\tmpbox}{Keywords:}
\newlength{\keyword}
\setlength{\keyword }{\textwidth}
\addtolength{\keyword}{-\wd\tmpbox}

\newpage
\section{Introduction}

     A correct description of both  the
     ground and excited states of scalar mesons encounters a
     variety of complex problems. Let us point out
     some of them. 
     i) For a long time,  the experimental
     status of the lightest scalar isoscalar singlet
     meson was unclear. In some papers, the resonance
     $f_0(1370)$ was considered as that particle
     \cite{F1370}, and it was not until 1998 that the resonance
     $f_0(400-1200)$ was included into the
     summary tables of PDG review%
     \footnote{ However, in earlier editions
     of PDG  the light $\sigma$ state still could be
     found;  it was excluded later.  }
     \cite{PDG}.
     ii) The scalar
     isoscalar states, as having  quantum numbers
     of vacuum, are most probably get mixed with
     glueballs  \cite{glueball}. 
     iii) There is also a
     lot of problems related to the description of
     $f_0(980)$ and $a_0(980)$. Their unusual
     experimental branching ratios for several decays
     have brought forth different ideas concerning the
     structure of the mesons. Among them, there are the
     quark-antiquark model \cite{F1370,glueball,volk}, the
     four-quark model \cite{achasov} and the kaon
     molecule model \cite{kmolecule}.  
     iv) The strange meson
     $K^*_0(1430)$ seems too heavy to be the ground state: $1\gev$
     is more characteristic of the ground meson states (see \cite{ishida}).

The description of the ground and excited states of the pion, kaon and the
vector meson nonet in the framework of a nonlocal version of the NJL model
has been done in our earlier papers \cite{weiss,volk97,ven}. Here we intend to
study the ground and first radially excited states of  $\eta$, $\eta'$ mesons
and of the scalar meson nonet.

To produce correct masses for
the ground states of $\eta$ and $\eta'$, we, as usual,  introduce
't~Hooft interaction \cite{klev,volk98}. Although our model is nonlocal,
which is
reflected in the presence of form factors in the four-quark vertices,
we, nevertheless, assume the 't~Hooft term local.
The form factors in  scalar channels of quark current-current
interaction are chosen identical to those in the pseudoscalar channel. This
is a requirement of the global chiral symmetry of quark interaction.
With that assumption, there  is no need for additional parameters in the 
form factors of scalar quark vertices.
Therefore, the masses of scalar mesons can be immediately predicted
after fixing the form factor parameters by the pseudoscalar meson masses from 
experiment. As a result, we have found that the model masses of the radial 
excitations of scalar isoscalar mesons are close to the 
experimentally observed 
$f_0(1370)$, $f_J(1710)$%
\footnote{
We assume hereafter $f_J(1710)$ is an isoscalar ($J=0$).
}%
, $a_0(1450)$ mesons. This allows us to interpret them
as the first radial excitations of
mesons $f_0(400-1200)$, $f_0(980)$ and $a_0(980)$.
As to the state $f_0(1500)$, we are inclined
to consider it as a glueball.
In our further works  we will take
into account possible mixing of the  $\bar qq$ scalar meson states with glueballs
\cite{glueball}.

Concerning the strange scalar $\kstar$, we think that the state with mass $1430\mev$
is much likely a radial excitation of a light and wide resonance with
mass about $960\mev$ (see \cite{ishida}). Further 
discussion on this problem is given in 
conclusion.

As to the ground states $a_0(980)$ and $f_0(980)$, the detailed discussion
on their internal structure and properties is beyond the scope of our paper.

Our paper is organized as follows. In Sec.~2, we introduce the 
chiral quark Lagrangian with nonlocal four-quark vertices and 
local 't Hooft interaction. In Sec.~3, we calculate the effective
Lagrangian for  isovector and strange mesons in the one-loop 
approximation. There we renormalize meson fields and transform
the free part of the Lagrangian to the  diagonal form
and obtain meson mass formulae. Section 4 is devoted to  isoscalar mesons
where we find masses and mixing coefficients. The model parameters 
are discussed in Sec.~5. In Sec.~6, we calculate the widths of
major strong decays of  excited states of 
$a_0$,  $\sigma$ and $f_0$ mesons. The results of our work and 
possible ways to improve the model are
discussed in Sec.~7. Some details of the calculations 
fulfilled in Sec.~4 and 6 are given in Appendices A and B.

\section{$U(3)\times U(3)$ chiral Lagrangian with excited meson states
and 't Hooft interaction}
We use a  nonlocal separable four-quark interaction
of a current-current form which admits nonlocal vertices (form
factors) in the quark currents, and a pure local six-quark 't Hooft
interaction
\cite{klev,volk98}:
\ba
     {\cal L}(\bar q, q) &=&
     \int\! d^4x\; \bar q(x)
     (i \partialslash -m^0) q(x)+
     {\cal L}^{(4)}_{\rm int}+
     {\cal L}^{(6)}_{\rm int},  \label{lag}\\
     {\cal L}^{(4)}_{\rm int} &=&
     \int\! d^4x\sum^{8}_{a=0}\sum^{N}_{i=1}
     \frac{G}{2}[j_{S,i}^a(x) j_{S,i}^a(x)+
     j_{P,i}^a(x) j_{P,i}^a(x)],\\
     {\cal L}^{(6)}_{\rm int}&=&-K \left[\det
     \left[\bar q (1+\gamma_5)q\right]+
     \det\left[\bar q (1-\gamma_5)q\right]
     \right].
\ea
Here, $m^0$ is the current quark mass matrix ($m_u^0\approx m_d^0$) and
$j^a_{S(P),i}$ denotes the scalar (pseudoscalar) quark currents
\be
     j^a_{S(P),i}(x)=
     \int\! d^4x_1 d^4x_2\; \bar q(x_1)
     F^a_{S(P),i }(x;x_1,x_2) q(x_2)
\ee
where $ F^a_{S(P),i}(x;x_1,x_2)$ are the scalar (pseudoscalar)
nonlocal quark vertices.
To describe the first radial excitations of mesons, we take
the form factors in momentum space as follows (see \cite{weiss,volk97,ven}),
\be
\begin{array}{ll}
     F_{S,j}^a({\bf k})=\lambda^a
     f^a_j,\quad & F_{P,j}^a=
     i\gamma_5 \lambda^a f^a_j
\end{array}
\ee
\be
     f^a_1\equiv 1,\quad f^a_2\equiv f_a({\bf k})=c_a(1+d_a {\bf k}^2),\label{fDef}
\ee
where
$\lambda^a$ are Gell--Mann matrices, 
$\lambda^0 = {\sqrt{2\over 3}}${\bf 1}, with {\bf 1} being the unit matrix.
Here, we consider the form factors in 
the rest frame of mesons
\footnote{The form factors depend on the transversal parts of
the relative momentum of quark-antiquark pairs $k_{\perp} =
k - \frac{k\cdot P}{P^2}P$, where $k$ and
$P$ are the relative and total momenta of a quark-antiquark pair,
respectively. Then, in the rest frame of mesons, ${\bf P}_{meson}$ = 0,
the transversal momentum is
$k_{\perp } = (0, {\vec k})$,
and we can define the form factors as depending on the 3-dimensional momentum ${\vec k}$ alone.
}%
.

The part of the Lagrangian (\ref{lag}), describing the ground states and
first radial excitations, can be rewritten in the following form
(see \cite{klev} and \cite{volk98}):
\ba
{\cal L}&=& \int\! d^4x \biggl\{
     \bar q(x) (i\partialslash-m^0) q(x) +
     \frac{G}{2}\sum_{a=0}^8 \left[\left(j_{S,2}^a\right)^2+
     \left(j_{P,2}^a\right)^2\right]+ \nonumber\\
&&   \frac12\sum_{a=1}^9\left[G^{(-)}_a
          \left(\bar q(x) \tau_a q(x)\right)^2+
     G^{(+)}_a\left(\bar q(x)i\gamma_5 \tau_a q(x)\right)^2\right]+
     \label{lagr}\\
&&   G_{us}^{(-)}(\bar q(x)\lambda_u q(x)) (\bar q(x)\lambda_s q(x)) +
     G_{us}^{(+)}(\bar q(x)i\gamma_5\lambda_u q(x))
          (\bar q(x)i\gamma_5\lambda_s q(x))\biggr\},\nonumber
\ea
where
\ba
     &&{\tau}_i={\lambda}_i ~~~ (i=1,...,7),~~~
     \tau_8 = \lambda_u = ({\sqrt 2}
     \lambda_0 + \lambda_8)/{\sqrt 3},\nonumber\\
&&   \tau_9 = \lambda_s = (-\lambda_0 +
     {\sqrt 2}\lambda_8)/{\sqrt 3}, \label{DefG} \\
&&   G_1^{(\pm)}=G_2^{(\pm)}=G_3^{(\pm)}=
     G \pm 4Km_sI_1(m_s), \nonumber \\
&&   G_4^{(\pm)}=G_5^{(\pm)}=G_6^{(\pm)}=
     G_7^{(\pm)}= G \pm 4Km_uI_1(m_u),
     \nonumber \\
&&   G_u^{(\pm)}= G \mp 4Km_sI_1(m_s), ~~~
     G_s^{(\pm)}= G, ~~~
     G_{us}^{(\pm)}= \pm 4{\sqrt 2}Km_uI_1(m_u).\nonumber
\ea
Here $m_u$ and $m_s$ are the constituent quark masses and $I_1(m_q)$
is the integral which for an arbitrary $n$ is defined as follows
\be
     I_n(m_q)={-i N_c\over (2\pi)^4}
     \int_{\Lambda_3}\!d^4 k 
     {1\over (m^2_q-k^2)^n} .
     \label{DefI}
\ee
The 3-dimensional cut-off $\Lambda_3$ in (\ref{DefI})  is implemented
to regularize the divergent integrals%
\footnote{For instance,
$I_1(m)=\frac{N_c m^2}{8\pi^2}\left.[x\sqrt{x^2+1}-\ln(x+\sqrt{x^2+1})]\right|_{x=\Lambda_3/m}$.
}%
.

\section{The masses of isovector and strange mesons
(ground and excited states)}

After bosonization, the part of  Lagrangian (\ref{lagr}),
describing  the isovector and strange mesons, takes the form
\ba
&&   {\cal L}(a_{0,1},K_0^*{}_{,1},\pi_1,K_1, a_{0,2},  K_0^*{}_{,2}, \pi_2, K_2)=
     -\frac{a_{0,1}^2}{2G_{a_0}}-\frac{{K_0^*{}_{,1}}^2}{G_{K_0^*}}-\frac{\pi_1^2}{2G_\pi}-
     \frac{K_1^2}{G_K}-\nonumber\\
&&   \frac{1}{2G} ( a_{0,2}^2+2(K_0^*{}_{,2})^2+
     \pi_2^2+2 K_2^2)-\nonumber\\
    &&   i N_c \Tr\ln\left[1+
     \frac{1}{i\partialslash-m}\sum_{a=1}^7\sum_{j=1}^2 \lambda_a\left[
     \sigma^a_j+i\gamma_5\varphi^a_j\right]f^a_j
     \right]
\label{bosLag}
\ea
where $m=\diag(m_u,m_d,m_s)$ is the matrix of constituent quark masses 
($m_u\approx m_d$),
$\sigma^a_j$ and $\phi^a_j$  are the scalar and pseudoscalar fields:
$\sum_{a=1}^{3}(\sigma^a_j)^2\equiv a_{0,j}^2=({a_{0,j}^0})^2+2a_{0,j}^+a_{0,j}^-$, 
$\sum_{a=4}^{7}(\sigma^a_j)^2\equiv 2{\kstar{}_{,j}}^2=
2(\bar{\kstar}_{,j})^0(\kstar{}_{,j})^0+2(\kstar{}_{,j})^+(\kstar{}_{,j})^-$,
$\sum_{a=1}^{3}(\varphi^a_j)^2\equiv \pi_j^2=({\pi_j^0})^2+2\pi_j^+\pi_j^-$,
$\sum_{a=4}^{7}(\varphi^a_j)^2\equiv 2K_j^2=2\bar K^0_j K^0_j+2\bar K^+_jK^-_j$.
As to the coupling constants $G_a$, they 
will be defined later (see Sect.~5 and (\ref{DefG})).

The free part of Lagrangian (\ref{bosLag}) has the following form 
\ba
     {\cal L}^{(2)}(\sigma,\varphi)=\frac12\sum_{i,j=1}^2\sum_{a=1}^7
     \left(\sigma^a_i K_{\sigma,ij}^a(P)\sigma^a_j+
     \varphi^a_i K_{\varphi,ij}^a(P)\varphi^a_j\right)
     \label{L2}
\ea
    where the coefficients $K^{a}_{\sigma(\varphi), ij}(P)$ are given below,
\ba
    &&  K_{\sigma(\varphi), ij}^{a}(P)=
     -\delta_{ij}\left[\frac{\delta_{i1}}{G_a^{(\mp)}}+\frac{\delta_{i2}}{G}\right]-\nonumber\\
    &&   i N_c \Tr \int_{\Lambda_3}\frac{d^4k}{(2\pi)^4}
     {1\over \kslash+\Pslash/2-m^a_q}
     r^{\sigma(\varphi)}f_i^a
     {1\over \kslash-\Pslash/2-m^a_{q'}}
      r^{\sigma(\varphi)}f_j^a,
     \label{K_full}
\ea
\be
     r^\sigma=1,\quad r^{\phi}=i\gamma_5, 
\ee
\be
     m_q^a = m_u~~(a = 1,...,7);\quad
     m_{q'}^a = m_u~~(a = 1,...,3);~~ m_{q'}^a = m_s~~ (a = 4,...,7),
     \label{m_q^a}
\ee
with $m_u$ and $m_s$ being the constituent quark masses and $f_j^a$ defined in
(\ref{fDef}).
Integral (\ref{K_full}) is evaluated by expanding in the
meson field momentum $P$. To order $P^2$, one obtains
\ba
     K_{\sigma(\varphi),11}^a(P)&=& 
     Z_{\sigma(\varphi),1}^a (P^2 -
     (m_q^a\pm m_{q'}^a)^2- M_{\sigma^a(\varphi^a),1}^2 ),\nonumber\\
     K_{\sigma(\varphi),22}^a(P)
	 &=& Z_{\sigma(\varphi), 2}^a (P^2 -
     (m_q^a\pm m_{q'}^a)^2-  M_{\sigma^a(\varphi^a),2}^2 ),
     \nonumber \\
     K_{\sigma(\varphi),12}^a(P) &=& K_{\sigma(\varphi),21}^a(P) \;\; = \;\;
     \gamma_{\sigma(\varphi)}^a (P^2 - (m^a_q\pm m^a_{q'})^2 ),
     \label{Ks_matrix}
\ea
where 
\be
     Z_{\sigma,1}^a = 4 I_2^a, \hspace{2em}  Z_{\sigma,2}^a  =  4 I_2^{ff a},
     \hspace{2em} \gamma_{\sigma}^a  = 4 I_2^{f a},
     \label{Zs} 
\ee
\be
     Z_{\varphi,1}^a = Z Z_{\sigma,1}^a, 
     \hspace{2em}  Z_{\varphi,2}^a  = Z_{\sigma,2}^a  ,
     \hspace{2em} \gamma_{\varphi}^a  = Z^{1/2}\gamma_{\sigma}^a
     \label{Zp} 
\ee
and
\ba
      M_{\sigma^a(\varphi^a),1}^2 &=& (Z_{\sigma(\varphi),1}^a)^{-1}
     \left[\frac{1}{G_a^{(\mp)}}-4(I_1(m_q^a) +
     I_1(m_{q'}^a))\right]
     \label{M_1} \\
     M_{\sigma^a(\varphi^a),2}^2 &=& (Z_{\sigma(\varphi),2}^a)^{-1}
     \left[\frac{1}{G}-4(I_1^{ff a}(m_q^a) +
     I_1^{ff a}(m_{q'}^a))\right].
     \label{M_2}
\ea
The factor $Z$ here appears due to account of $\pi-a_1$-transitions~\cite{volk,volk97},
\be
Z=1-\frac{6 m_u^2}{ M_{a_1}^2},
\ee
and the integrals $I_2^{f..f}$  contain form factors:
\be
     I_2^{f..f_a}(m^a_q,m^a_{q'})={-i N_c\over (2\pi)^4}
     \int_{\Lambda_3} d^4 k 
     {f_a({\bf k})..f_a({\bf k})\over ((m^a_q)^2-k^2)((m^a_{q'})^2-k^2)}.
     \label{DefIf}
\ee
Further, we consider only the scalar isovector and strange 
mesons because the masses of the
pseudoscalar mesons have been already described in \cite{volk97}.

After the renormalization of the scalar fields
\be
     \sigma_i^{a r}=\sqrt{Z_{\sigma,i}^a} \sigma_i^{a}  \label{renorm}
\ee
the part of Lagrangian (\ref{L2}) which describes the scalar mesons
takes the form
\ba
     {\cal L}^{(2)}_{a_0}&=&\frac12
     \left(
     P^2-4 m^2_u -M^2_{a_0, 1}\right)a_{0, 1}^2+ \Gamma_{a_0}\left(
     P^2-4m_u^2\right)a_{0, 1}a_{0, 2}+\nonumber\\
     &&\frac12\left(P^2-4m_u^2- M_{a_{0},2}^2\right)a_{0, 2}^2,
     \label{La0}
\ea
\ba
     {\cal L}^{(2)}_{\kstar}\!&=&\!\frac12\! \left(
     P^2-(m_u+ m_s)^2\! -\! M^2_{\kstar,1}\right)\kstar{}_{,1}^2\!
     +\! \Gamma_{\kstar}
     \left(P^2-(m_u+m_s)^2\right)\kstar{}_{,1}\kstar{}_{,2}+\nonumber\\
     &&\!\frac12\left(P^2-(m_u+m_s)^2- M_{\kstar{},2}^2\right)\kstar{}_{,2}^2,
     \label{LK}
\ea
where
\be
     \Gamma_{\sigma^a}=\frac{I_2^{f_a}}{\sqrt{I_2 I_2^{ff_a}}}.\qquad
\ee
After the transformations of the meson fields
\ba
     \sigma^a
     &=& \cos( \theta_{\sigma,a} - \theta_{\sigma,a}^0) \sigma_1^{ar}
     - \cos( \theta_{\sigma,a} + \theta_{\sigma,a}^0) \sigma_2^{ar},   \nonumber \\
     \hat\sigma^{a}
     &=& \sin ( \theta_{\sigma,a} - \theta_{\sigma,a}^0) \sigma_1^{ar}
     - \sin ( \theta_{\sigma,a} + \theta_{\sigma,a}^0) \sigma_2^{ar},
     \label{transf}
\ea
Lagrangians (\ref{La0}) and (\ref{LK}) take the diagonal form:
\ba
     L_{a_0}^{(2)} &=& \half (P^2 - M_{a_0}^2)~ a_0^2 +
     \half (P^2 - M_{\hat a_0}^2)\hat a_0^{ 2}, \\
     L_{\kstar}^{(2)} &=& \half (P^2 - M_{\kstar}^2)~ \kstar{}^2 +
     \half (P^2 - M_{\hat\kstar}^2)\hat \kstar{}^{ 2}.
     \label{L_pK}
\ea
Here we have
\ba
&&   M^2_{(a_0, \hat a_0)} = \frac{1}{2 (1 - \Gamma^2_{a_0})}
     \biggl[M^2_{a_{0}, 1} + M^2_{a_{0}, 2}\pm \nonumber \\
&&   \qquad \sqrt{(M^2_{a_{0}, 1} - M^2_{a_{0}, 2})^2 +
     (2 M_{a_{0}, 1} M_{a_{0}, 2} \Gamma_{a_0})^2}\biggr]+4m_u^2, \\
&&   M^2_{(\kstar, \hat\kstar)} = \frac{1}{2 (1 - \Gamma^2_{\kstar})}
     \biggl[M^2_{\kstar, 1} + M^2_{\kstar, 2}\pm
       \nonumber \\
&&   \qquad  \sqrt{(M^2_{\kstar,1} - M^2_{\kstar,2})^2 +
     (2 M_{\kstar,1} M_{\kstar,2} \Gamma_{\kstar})^2}\biggr]+ (m_u+m_s)^2,
     \label{MpKstar}
\ea   
and 

\be
     \tan 2 {\bar{\theta}}_{\sigma,a} = \sqrt{\frac{1}{\Gamma_{\sigma^a}^2} -
     1}~\left[ \frac{M_{\sigma^a,1}^2-  M_{\sigma^a,2}^2}{M_{\sigma^a,1}^2
     + M_{\sigma^a,2}^2} \right],\qquad
     2 \theta_{\sigma,a} = 2 {\bar{\theta}}_{\sigma,a} + \pi,
     \label{tan}
\ee
\be
     \sin \theta_{\sigma,a}^{0} =\sqrt{{1+\Gamma_{\sigma^a}}\over 2}.
     \label{theta0}
\ee
The caret symbol stands for the first radial excitations of mesons.
Transformations (\ref{transf}) express the ``physical'' fields $\sigma$ and
$\hat\sigma$ through the ``bare'' ones $\sigma^{ar}_i$ and for calculations,
these equations must be inverted. 
For practical use, we collect 
 the values of the inverted equations for the scalar and
pseudoscalar fields%
\footnote{
Although the formulae for the pseudoscalars are not displayed here  
(they  have been already obtained in \cite{volk97}) we need the
values because we are going to calculate the decay widths of processes
where pions and kaons are  secondary particles.
}
 in Table \ref{mixingTable}.
\begin{table}
\caption{The mixing coefficients for the ground and first radially excited states of 
the scalar and pseudoscalar isovector and strange mesons. 
The caret symbol marks the excited states.}
\label{mixingTable}
$$
	\begin{array}{||r|c|c||}
	\hline
		& a_0 & \hat a_0\\	
	\hline
	a_{0,1}	&0.87	  &0.82    \\
	a_{0,2} &0.22	  &-1.17   \\
	\hline
	\end{array}
\quad
	\begin{array}{||r|c|c||}
	\hline
		& \kstar & \hat\kstar\\	
	\hline
	\kstar{}_{,1}	&0.83	  &0.89    \\
	\kstar{}_{,2} &0.28	  &-1.11   \\
	\hline
	\end{array}
$$
$$
	\begin{array}{||r|c|c||}
	\hline
		& \pi & \hat\pi\\	
	\hline
	\pi_1	&1.00	  &0.54    \\
	\pi_2   &0.01	  &-1.14   \\
	\hline
	\end{array}
\quad
	\begin{array}{||r|c|c||}
	\hline
		& K & \hat K   \\	
	\hline
	K_1	&0.96	  &0.56   \\
	K_2    & 0.09	  &-1.11  \\
	\hline
	\end{array}
$$

\end{table}


\section{The masses of isoscalar mesons (the ground and excited states)}

The 't Hooft interaction effectively gives rise to the additional
four-quark vertices in the isoscalar part of Lagrangian (\ref{lagr}):
\be
     {\cal L}_{\rm isosc}=\sum_{a,b=8}^9\left[
     (\bar q \tau_a q)T^S_{a b}
     (\bar q \tau_b q)+
     (\bar q i\gamma_5\tau_a q)T^P_{a b}
     (\bar q i\gamma_5\tau_b q)\right]
\ee
where $T^{S(P)}$ is a matrix with elements defined as follows
(for the definition of $G_u^{(\mp)}$, $G_s^{(\mp)}$ and $G_{us}^{(\mp)}$ see (\ref{DefG}))
\be
      \begin{array}{ll}
     T^{S(P)}_{88}=G^{(\mp)}_{u}/2,\quad & T^{S(P)}_{89}=G^{(\mp)}_{us}/2, \\ 
     T^{S(P)}_{98}=G^{(\mp)}_{us}/2,\quad & T^{S(P)}_{99}=G^{(\mp)}_{s}/2.
     \end{array}
\ee
This leads
to nondiagonal terms in the free part of the
effective Lagrangian for isoscalar
scalar and pseudoscalar mesons after bosonization
\ba
     &&{\cal L}_{\rm isosc}(\sigma,\varphi)=
	-{1\over 4}\sum_{a,b=8}^9\left[
     \sigma^{a}_1
     (T^S)^{-1}_{a b}\sigma^{b}_1 + 
     \varphi^{a}_1
     (T^P)^{-1}_{a b}\varphi^{b}_1\right]-
\nonumber \\
     &&  {1\over 2G}\sum_{a=8}^9 \left[\left(\sigma^{a}_2\right)^2 + 
     \left(\varphi^{a}_2\right)^2
     \right]-
\nonumber \\
     &&i~{\rm Tr}\ln \left\{1 + {1\over i{ \partialslash} - m}
	\sum_{a=8}^9\sum_{j=1}^2
	\tau^{a}[
     \sigma^{a}_j +
     i\gamma_5 
     \varphi^{a}_j
     ]f^{a}_j \right\},  \label{Lbar}
\ea
where $(T^{S(P)})^{-1}$ is the inverse of $T^{S(P)}$:
\be
 \begin{array}{ll}
     (T^{S(P)})^{-1}_{88}=2G^{(\mp)}_{s}/D^{(\mp)},\quad & 
	(T^{S(P)})^{-1}_{89}= (T^{S(P)})^{-1}_{98}=-2G^{(\mp)}_{us}/D^{(\mp)}, \\ 
	(T^{S(P)})^{-1}_{99}=2G^{(\mp)}_{u}/D^{(\mp)},\quad &
	D^{(\mp)}=G^{(\mp)}_u G^{(\mp)}_s-(G^{(\mp)}_{us})^2 .
 \end{array}
     \label{Tps1}
\ee
From (\ref{Lbar}), in the one-loop approximation, one obtains the
free part of the effective Lagrangian
\ba
     {\cal L}^{(2)}(\sigma,\phi)=\frac12\sum_{i,j=1}^2\sum_{a,b=8}^9
     \left(\sigma^a_i K_{\sigma,ij}^{[a,b]}(P)\sigma^b_j+
     \varphi^a_i K_{\phi,ij}^{[a,b]}(P)\varphi^b_j\right).
	\label{Lisosc}
\ea
The definition of $K_{\sigma(\varphi),i}^{[a,b]}$ is given in 
Appendix A.

After  the renormalization of both the scalar and pseudoscalar fields,
 analogous to (\ref{renorm}),  we come to the Lagrangian which
can be represented in a form slightly different from that
of (\ref{Lisosc}). It is convenient to
introduce 4-vectors of ``bare'' fields
\be
     \Phi=(\varphi_{1}^{8\,r},\varphi_{2}^{8\,r},
	\varphi_{1}^{9\,r},\varphi_{2}^{9\,r}), \qquad
     \Sigma=(\sigma_{1}^{8\,r},\sigma_{2}^{8\,r},
	\sigma_{1}^{9\,r},\sigma_{2}^{9\,r}).
\ee
Thus, we have
\ba
     {\cal L}^{(2)}(\Sigma,\Phi)=\frac12\sum_{i,j=1}^4
     \left(\Sigma_i {\cal K}_{\Sigma,ij}(P)\Sigma_j+
     \Phi_i {\cal K}_{\Phi,ij}(P)\Phi_j\right)
     \label{L2a}
\ea
where we introduced new functions ${\cal K}_{\Sigma(\Phi),ij}(P)$ (see Appendix A).

Up to this moment one has four pseudoscalar and four scalar meson states
which are the octet and nonet singlets. The mesons of the same parity
have the same quantum numbers and, therefore,
they are expected to be mixed. In our model the mixing is represented
by $4\times 4$ matrices $R^{\sigma(\varphi)}$ which
transform the ``bare'' fields $\varphi_{i}^{8\,r}$, $\varphi_{i}^{9\,r}$,
$\sigma_{i}^{8\,r}$ and  $\sigma_{i}^{9\,r}$
entering the 4-vectors $\Phi$ and $\Sigma$  
to the ``physical'' ones $\eta$,  $\eta'$,  $\hat\eta$,  $\hat\eta'$,
 $\sigma$, $\hat\sigma$, $f_0$ and $\hat f_0$ represented as components
of vectors
$\Phi_{\rm ph}$ and $\Sigma_{\rm ph}$:
\be
     \Phi_{\rm ph}=(\eta,\hat\eta,\eta',\hat\eta'), \qquad
     \Sigma_{\rm ph}=(\sigma,\hat\sigma,f_0,\hat f_0)
\ee
where, let us remind once more,  a caret over a meson field 
stands for the first radial excitation
of the meson. The transformation $R^{\sigma(\varphi)}$ is linear and nonorthogonal:
\be
     \Phi_{\rm ph}=R^{\varphi}\Phi,\qquad \Sigma_{\rm ph}=R^{\sigma}\Sigma.
\ee
In terms of ``physical'' fields the free part of the effective
Lagrangian is of the conventional form and the coefficients
of matrices $R^{\sigma(\varphi)}$ give the mixing of
the $\bar uu$ and $\bar ss$ components, with and without form factors.

Because of the complexity of the procedure of diagonalization for the
matrices of dimensions greater than 2,
there is no such simple formulae as, {\it e.g.}, in (\ref{transf}).
Hence, we do not implement it analytically but
use numerical methods to obtain matrix elements
(see Table~\ref{isoscMixTab}).
\begin{table}
\caption{The mixing coefficients for the isoscalar meson states}
\label{isoscMixTab}
$$
	\begin{array}{||r|c|c|c|c||}
	\hline\hline
	 		&\eta 		&\hat\eta 	&\eta' 		&\hat\eta'\\
	\hline
	\varphi^8_{1}	&0.71		&0.62		&-0.32		&0.56		\\
	\varphi^8_{2}   &0.11		&-0.87		&-0.48		&-0.54		\\
	\varphi^9_{1}	&0.62		&0.19		&0.56		&-0.67		\\
	\varphi^9_{2}   &0.06		&-0.66		&0.30		&0.82		\\
	\hline
	\end{array}
$$
$$
	\begin{array}{||r|c|c|c|c||}
	\hline\hline
	 		&\sigma 	&\hat\sigma 	&f_0 		&\hat f_0\\
	\hline
	\sigma^8_{1}	&-0.98		&-0.66		&0.10		&0.17		\\
	\sigma^8_{2}&0.02		&1.15		&0.26		&-0.17		\\
	\sigma^9_{1}	&0.27		&-0.09		&0.82		&0.71		\\
	\sigma^9_{2}&-0.03		&-0.21		&0.22		&-1.08		\\
	\hline
	\end{array}
$$
\end{table}
%

\section{Model parameters and meson masses}

In our model we have five basic parameters: the masses of
the constituent $u(d)$ and $s$ quarks, $m_u=m_d$ and
$m_s$, the cut-off parameter $\Lambda_3$, the four-quark
coupling constant $G$ and the  't Hooft coupling constant
$K$. We have fixed these parameters with the help  of 
input parameters: the pion decay constant $F_\pi=93 \mev$,
the $\rho$-meson decay constant $g_\rho=6.14$
(decay $\rho\to2\pi$)%
\footnote{Here we do not consider vector and axial-vector mesons,
however, we have used  the relation $g_\rho=\sqrt{6}g_{\sigma}$
 together with the Goldberger--Treiman relation
$g_\pi=\frac{m}{F_\pi}=Z^{-1/2}g_{\sigma}$ to fix the parameters
$m_u$ and $\Lambda_3$ (see \cite{volk97}).
},
the masses of pion and kaon and the mass difference of $\eta$ and $\eta'$
mesons (for details of these calculations, see \cite{volk97,ven,volk98}).
Here we give only numerical estimates of these parameters:
\ba
	&&m_u=280\mev,\quad m_s=405 \mev, \quad\Lambda_3=1.03 \gev, \nonumber\\
	&&G=3.14\gev^{-2},\quad K=6.1\gev^{-5}.
\ea
We also have a set of additional parameters $c^{\sigma^a(\varphi^a)}_{qq}$
in form factors $f^a_2$. These parameters are defined by masses of
excited pseudoscalar mesons,
 $c_{uu}^{\pi,a_0}=1.44$, $c_{uu}^{\eta,\eta',\sigma,f_0}=1.5$,
$c_{us}^{K,\kstar}=1.59$,
$c_{ss}^{\eta,\eta',\sigma,f_0}=1.66$.
The slope parameters $d_{qq}$ are fixed by special conditions satisfying 
the standard gap equation,
$d_{uu}=-1.78 \gev^{-2}$,
$d_{us}=-1.76 \gev^{-2}$,  $d_{ss}=-1.73 \gev^{-2}$ (see \cite{volk97}). 
Using these parmeters, we obtain masses of pseudoscalar and scalar mesons which
are listed in Table \ref{masses} together with experimental values.
\begin{table}
\caption{The model masses of mesons, MeV}
\label{masses}
$$
\begin{array}{||l|c|c|c|c||}
	\hline \hline
		& GR 		& EXC		& GR(Exp.)		&EXC(Exp.) 	\\
	\hline
M_{\sigma} 	& 530		& 1330		&400-1200		&1200-1500	\\
M_{f_0}		& 1070		& 1600		&980\pm10		&1712\pm5	\\
M_{a_0}		& 830		& 1500		&983.4\pm0.9		&1474\pm19	\\
M_{\pi} 	& 140		& 1300		&139.56995\pm0.00035	&1300\pm100	\\
M_{K} 		& 490		& 1300		&497.672\pm0.031	&1460(?)	\\
M_{\kstar} 	& 960		& 1500		&-	&1429\pm12	\\
M_{\eta} 	& 520		& 1280		&547.30\pm0.12		&1297.8\pm2.8	\\ 
M_{\eta'} 	& 910		& 1470		&957.78\pm0.14		&1440-1470	\\
	\hline \hline
\end{array}
$$
\end{table}

From our calculations we come to the following interpretation of
$f_0(1370)$, $f_J(1710)$ and $a_0(1470)$ mesons: we consider them
as the first radial excitations of the ground states 
$f_0(400-1200)$, $f_0(980)$ and $a_0(980)$. Meanwhile,
the meson $f_0(1500)$ is much likely a glueball. The strong
decays which we consider in the next section 
 substantiate our point of view.

\section{Strong decays of the scalar mesons}
The ground and excited states of scalar mesons $f_0$, $a_0$ 
decay mostly into  pairs of pseudoscalar
mesons. In the framework of a quark model and in the leading order
of $1/N_c$ expansion, the
processes are described by triangle quark
diagrams (see Fig.1).
Before we start to calculate the amplitudes, corresponding to
these diagrams,
we introduce, for convenience, Yukawa coupling constants 
which naturally appear after
the renormalization (\ref{renorm}) of  meson fields:
\ba
	&&g_{\sigma_u}\equiv \left.g_{\sigma^a}\right|_{a=1,2,3,8}=[4I_2(m_u)]^{-1/2},
	\quad
	g_{\kstar}\equiv \left.g_{\sigma^a}\right|_{a=4,5,6,7}=[4I_2(m_u,m_s)]^{-1/2},
	\nonumber\\
	&&g_{\sigma_s}\equiv g_{\sigma^9}=[4I_2(m_s)]^{-1/2},
	\quad
	g_{\varphi^a}=Z^{-1/2}g_{\sigma^a}
	\nonumber\\
	&&g_{\pi}\equiv\left. g_{\varphi^a}\right|_{a=1,2,3},
	\quad
	g_{K}\equiv \left.g_{\varphi^a}\right|_{a=4,5,6,7},
	\quad
	g_{\varphi_u}\equiv g_{\varphi^8},
	\quad 
	g_{\varphi_s}\equiv g_{\varphi^9}
     \label{g}
\ea
\ba
	&&\hat g_{\sigma_u}\equiv \left.\hat g_{\sigma^a}\right|_{a=1,2,3,8}=[4I_2^{ff}(m_u)]^{-1/2},
	\;\,
	\hat g_{\kstar}\equiv \left.\hat g_{\sigma^a}\right|_{a=4,5,6,7}=[4I_2^{ff}(m_u,m_s)]^{-1/2},
	\nonumber\\ 
	&&\hat g_{\sigma_s}\equiv \hat g_{\sigma^9}=[4I_2^{ff}(m_s)]^{-1/2},
	\quad
	\hat g_{\varphi^a}=\hat g_{\sigma^a}
	\nonumber\\
	&&\hat g_{\pi}\equiv\left.\hat g_{\varphi^a}\right|_{a=1,2,3},
	\;\,
	\hat g_{K}\equiv \left.\hat g_{\varphi^a}\right|_{a=4,5,6,7},
	\;\,
	\hat g_{\varphi_u}\equiv \hat g_{\varphi^8},
	\;\, 
	\hat g_{\varphi_s}\equiv \hat g_{\varphi^9}
     \label{gX}
\ea

They can easily be related to $Z^a_{\sigma(\varphi),i}$ introduced in the
beginning of our paper.
Thus, the one-loop contribution to the effective Lagrangian can be 
rewritten in terms of the renormalized fields:
\ba
  {\cal L}_{\rm 1-loop}(\sigma,\varphi)&=&
      i N_c \Tr\ln\left[1+
     \frac{1}{i\partialslash-m}\sum_{a=1}^9 \tau_a\left[
     g_{\sigma^a}\sigma^a_1+i\gamma_5 g_{\varphi^a}\varphi^a_1+\right.\right.\nonumber\\
	&&\left.(\hat g_{\sigma^a}\sigma^a_2+i\gamma_5 \hat g_{\varphi^a}\varphi^a_2)f_a\right]
     \Biggr]
\label{bosLag1}
\ea

All amplitudes that describe processes of the type $\sigma\to\varphi_1\varphi_2$
can be divided into two parts:
\ba
	T_{\sigma\to\varphi_1\varphi_2}&=&
	C\left(-\frac{i N_c}{(2\pi)^4}\right)
	\int_{\Lambda_3}d^4 k \frac{\Tr[(m+\Slash{k}+\Slash{p}_1)\gamma_5
	(m+\Slash{k})\gamma_5(m+\Slash{k}-\Slash{p}_2)]}{
	(m^2-k^2)(m^2-(k+p_1)^2)(m^2-(k-p_2)^2)}\nonumber\\
	&=& 4mC\left(-\frac{i N_c}{(2\pi)^4}\right)
	\int_{\Lambda_3}d^4k
	\frac{\left[1-\frac{p_1\cdot p_2}{m^2-k^2}\right]}{(m^2-(k+p_1)^2)(m^2-(k-p_2)^2)}
	\nonumber\\
	&=&4 m C [I_2(m,p_1,p_2)-p_1\cdot p_2 I_3(m,p_1,p_2)]=T^{(1)}+T^{(2)}
	\label{T}
\ea
here $C=4 g_{\sigma} g_{\varphi_1}g_{\varphi_2}$ and $p_1$, $p_2$ are
momenta of the pseudoscalar mesons. 
Using (\ref{g}) and (\ref{gX}), we rewrite the amplitude
$T_{\sigma\to\varphi_1\varphi_2}$
in another form
\ba
&&T_{\sigma\to\varphi_1\varphi_2}\approx
4mZ^{-1/2}g_{\varphi_1} \left[1-p_1\cdot p_2
\frac{I_3(m)}{I_2(m)}\right],\label{T1}\\
&&p_1\cdot p_2 =\frac12(M_\sigma^2-M_{\varphi_1}^2-M_{\varphi_2}^2).
\ea
We assumed here that the ratio of $I_3$ to $I_2$  slowly changes with momentum
in comparison with factor $p_1\cdot p_2$,
therefore, we ignore their momentum dependence in (\ref{T1}). 
With this assumption we
are going to obtain just a qualitative picture for  decays of the
excited scalar mesons.

In eqs. (\ref{T}) and (\ref{T1}) we omitted the contributions
from the diagrams which include form factors in vertices.
The whole set of diagrams consists of those containing zero, one, two
and three form factors. To obtain the complete amplitude, one must
sum up all contributions.

After these general comments, let us  consider the decays of
$ a_0(1450)$, $f_0(1370)$ and $f_J(1710)$.
First, we estimate the
decay width of the process $\hat a_0\to\eta\pi$,
taking the mixing coefficients from Table~\ref{mixingTable} and~\ref{isoscMixTab}
(see Appendix B for the details).
The result is
\be
      T^{(1)}_{\hat a_0\to\eta\pi}\approx0.2\gev,
\ee
\be
     T^{(2)}_{\hat a_0\to\eta\pi}\approx3.5 \gev,
\ee
\be
     \Gamma_{\hat a_0\to\eta\pi}\approx
     160 \mev.
\ee

From this calculation one can see that $T^{(1)}\ll T^{(2)}$ and the
amplitude is dominated by its second part, $T^{(2)}$, which is
momentum dependent. The first part is small because the diagrams
with different numbers of form factors cancel each other. As a consequence,
in all processes where an excited scalar meson decays into a pair of
ground pseudoscalar states, the second part of the amplitude
defines the rate of the process.

For the decay $\hat a_0\to\pi\eta'$ we obtain the amplitudes
\be
     T^{(1)}_{\hat a_0\to\pi\eta'}\approx0.8 \gev,
\ee
\be
     T^{(2)}_{\hat a_0\to\pi\eta'}\approx3 \gev,
\ee
and the decay width
\be
    \Gamma_{\hat a_0\to\pi\eta'}\approx36 \mev.
\ee
The decay of $\hat a_0$ into kaons is described by the amplitudes $T_{\hat a_0\to K^+K^-}$
and  $T_{\hat a_0\to \bar K^0K^0}$
which, in accordance with our scheme,  can again be  divided into two parts: $T^{(1)}$
and $T^{(2)}$ (see Appendix B for details): 
\be
T_{\hat a_0\to K^+K^-}^{(1)}\approx 0.2\gev,
\ee
\be
T_{\hat a_0\to K^+K^-}^{(2)}\approx 2.1\gev.
\ee
and the  decay width  is
\be
\Gamma_{\hat a_0\to KK}=\Gamma_{\hat a_0\to K^+K^-}+\Gamma_{\hat a_0\to \bar K^0K^0}\approx 100\mev.
\ee
Qualitatively, our results do not contradict the experimental data.
\be
	\Gamma^{\rm tot}_{\hat a_0}=265\pm13 \mev,\quad BR(\hat a_0\to KK):BR(\hat a_0\to\pi\eta)= 0.88\pm0.23.
\ee
The decay widths
of radial excitations of scalar isoscalar mesons 
are estimated in the same way as it was shown above. We obtain:
\be
\Gamma_{\hat\sigma\to\pi\pi}=\left\{
\begin{array}{l}
550 \mev (M_{\hat\sigma}=1.3 \gev) \\
460 \mev (M_{\hat\sigma}=1.25 \gev),
\end{array}
\right.
\ee
\be
\Gamma_{\hat\sigma\to\eta\eta}=\left\{
\begin{array}{l}
24 \mev (M_{\sigma}=1.3 \gev) \\
15 \mev (M_{\sigma}=1.25 \gev),
\end{array}
\right.
\ee
\be
\Gamma_{\hat\sigma\to\sigma\sigma}=\left\{
\begin{array}{l}
6 \mev (M_{\sigma}=1.3 \gev) \\
5 \mev (M_{\sigma}=1.25 \gev),
\end{array}
\right.
\ee
\be
\Gamma_{\hat\sigma\to KK}\sim 5 \mev,
\ee
\be
\begin{array}{lclclcl}
\Gamma_{f_0(1710)\to 2\pi}& =& 3\mev, &\quad &
     \Gamma_{f_0(1500)\to 2\pi}&=& 3\mev,\\
\Gamma_{f_0(1710)\to 2\eta}& =& 40\mev, &\quad &
     \Gamma_{f_0(1500)\to 2\eta}&=& 20\mev,\\
\Gamma_{f_0(1710)\to \eta\eta'}& =& 42\mev, &\quad &
     \Gamma_{f_0(1500)\to \eta\eta'}&=& 10\mev,\\
\Gamma_{f_0(1710)\to KK}& =& 24\mev, &\quad &
     \Gamma_{f_0(1500)\to KK}&=& 20\mev.
\end{array}
\ee
The decays of $f_0(1500)$ and $f_0(1710)$ to $\sigma\sigma$ are negligibly small,
so we disregard them.

Here we desplayed our estimates  for both $f_J(1710)$ and $f_0(1500)$
resonances. Comparing them will allow us to decide which one to 
consider as the first radial excitation of $f_0(980)$ and which a glueball.
From the experimental data:
\be
\Gamma^{\rm tot}_{\sigma'}=200 - 500 \mev,
\quad
\Gamma^{\rm tot}_{f_0(1710)}=133\pm 14 \mev,
\quad
\Gamma^{\rm tot}_{f_0(1500)}=112\pm 10 \mev
\ee
we can see that in the case of $f_0(1500)$ being a $\bar qq$ state
there is deficit in the decay widths whereas for $f_J(1710)$ the result is
close to experiment.
From this we conclude that  the meson
$f_J(1710)$ is a radially excited partner for $f_0(980)$
and the meson state $f_0(1370)$ is the
first radial excitation  of $f_0(400-1200)$.
As to the state $f_0(1500)$, it is mostly a glueball which 
significantly contributes to the decay width.

For the decay widths of ground scalar states  the situation
is opposite to that for excited states. Indeed, the parts $T^{(1)}$
and $T^{(2)}$ of the amplitude $T$ are of the same order of magnitude 
and different in
sign, thus,  cancelling each other. The values for
decay widths of processes $\sigma\to\pi\pi$, $f_0(980)\to\pi\pi$
and $a_0(980)\to\pi\eta$ turn out to be too small in our approximation:
\be
\Gamma_{\sigma\to\pi\pi}\sim 350 \mev,\quad
\Gamma_{f_0(980)\to\pi\pi}\sim 25\mev,\quad
\Gamma_{a_0(980)\to\pi\eta}\sim 1\mev,
\ee
whereas the experiment gives us
\ba
&&\Gamma_{\sigma\to\pi\pi}\sim 600-1000 \mev,\quad
\Gamma_{f_0(980)\to\pi\pi}\sim 40-100\mev,\nonumber\\
&&\Gamma_{a_0(980)\to\pi\eta}\sim 50-100\mev.
\ea
Therefore, to describe correctly
strong decays of ground states of scalar mesons,
it is necessary to calculate the parts $T^{(1)}$
of the amplitudes more accurately, taking into account
their momentum dependence.
Indeed, from our previous paper \cite{conf} we know that 
taking into account the quark confinement and the momentum 
dependence of $I_2(m,p_1,p_2)$ one can find that
$\sigma$ decays into $\pi\pi$ with the width
\be
	\Gamma_{\sigma\to\pi\pi}\approx 700\mev.
\ee
For the decay $a_0\to\eta\pi$, the result can increase more
than by an order. 
A proper description of the
decay $a_0(980)\to\pi\eta$ can possibly be obtained from  the four-quark
interpretation of $a_0(980)$ state \cite{achasov}, and the four-quark 
component may dominate the process; 
however, this  is beyond our model.
For a more careful description of the decay $f_0(980)\to\pi\pi$ one should
take into account also the mixing with the glueball state $f_0(1500)$, which
we are going to do in our further work.

\section{Discussion and Conclusion}
Our calculations have shown 
that we can interpret the scalar states $f_0(1370)$, $a_0(1450)$ and $f_0(1710)$
as the first radial excitations of $f_0(400-1200)$, $a_0(980)$ and $f_0(980)$. 
We estimated their masses and the widths of main decays in the framework of a nonlocal
chiral quark model.
We would like to emphasize that we have not used
  additional parameters except those necessary to fix the
mass spectrum of pseudoscalar mesons. We used the same form factors both
for the scalar and pseudoscalar mesons, which is a requirement of the global chiral symmetry.

We assumed that the state $f_0(1500)$ is  a glueball, and its probable mixing
with $f_0(980)$, $f_0(1370)$ and $f_J(1710)$ may provide us with a 
more correct description of the masses of these states%
\footnote{
Our estimates for the masses of $f_0$ and $\hat f_0$: 
$M_{f_0}=1070\mev$ and $M_{\hat f_0}=1600\mev$ are expected to shift
to $M_{f_0}=980\mev$ and $M_{\hat f_0}=1710\mev$ after mixing with
the glueball $f_0(1500)$.
}%
(see Table~\ref{masses}).  We are going to consider this problem in a 
subsequent publication.

More complicated situation takes place for the ground state $a_0(980)$.
In the framework of our quark-antiquark model, we have a mass deficit for this meson,
$830 \mev$ instead of $980\mev$ and a small decay width for $a_0\to\pi\eta$.
We suspect that this drawback is caused by  four-quark
component in this state which we did not take into account \cite{achasov}.

There is also some mystery about the strange scalar meson $\kstar$. Its experimental 
mass is large enough, $M_{\kstar}=1430\mev$ and the width is $287\pm23\mev$.
In our model, there are two strange scalars, $\kstar(930)$ 
with a large width
and $\kstar(1500)$ with the width of the decay $\Gamma_{\kstar\to K\pi}\sim300\mev$. 
Thus, together with \cite{ishida} we suppose it is possible for a wide
strange resonance, $\kstar(930)$ to
exist in nature still missed in detectors as the ground state whereas
the resonance  
$\kstar(1430)$ is its radial excitation mostly decaying into $K\pi$. 
Our model gives for the excited meson $\kstar$: $M_{\hat\kstar}\approx 1500\mev$
and $\Gamma_{\hat\kstar\to K\pi}\approx 300\mev$.

In future we are going to take into account the presence of glueball states and
to develop a model with quark confinement for the description of heavy mesons.
We will also consider the decays of the excited $\eta$ and $\eta'$ mesons.

\section*{Acknowledgment}
We are very grateful to Prof. S.B.~Gerasimov for useful discussion.
This work has been supported by RFBR Grant N 98-02-16135 and
by Heisenberg--Landau Program 1999. 

\appendix

\section*{Appendix}

\section{Coefficients of the free part of effective Lagrangian
for the scalar isoscalar mesons.}

The functions $K_{\sigma(\varphi),ij}^{[a,a]}$ introduced in Sec.~4 (\ref{Lisosc})
are defined as follows
\ba
     K_{\sigma(\varphi),11}^{[a,a]}(P)&=& 
     Z_{\sigma(\varphi),1}^a (P^2 -
     (m_q^a\pm m_{q'}^a)^2-  M_{\sigma^a(\varphi^a),1}^2 ),\nonumber\\
     K_{\sigma(\varphi),22}^{[a,a]}(P)
	 &=& Z_{\sigma(\varphi), 2}^a (P^2 -
     (m_q^a\pm m_{q'}^a)^2-  M_{\sigma^a(\varphi^a),2}^2 ),
     \nonumber \\
     K_{\sigma(\varphi),12}^{[a,a]}(P) &=& K_{\sigma(\varphi),21}^{[a,a]}(P) \;\; = \;\;
     \gamma_{\sigma(\varphi)}^a (P^2 - (m_q^a\pm m_{q'}^a)^2 ),\\
     K_{\sigma(\varphi),11}^{[8,9]}(P)&=&K_{\sigma(\varphi),11}^{[9,8]}(P)=
	\frac12 \left(T^{S(P)}\right)^{-1}_{89},\nonumber\\
	K_{\sigma(\varphi),12}^{[8,9]}(P)&=&
	K_{\sigma(\varphi),12}^{[9,8]}(P)=
	K_{\sigma(\varphi),21}^{[8,9]}(P)=0,\nonumber\\
	K_{\sigma(\varphi),21}^{[9,8]}(P)&=&
	K_{\sigma(\varphi),22}^{[8,9]}(P)=
	K_{\sigma(\varphi),22}^{[9,8]}(P)=0. \nonumber
\ea
where the ``bare'' meson masses are
\ba
     && M_{\sigma^8(\varphi^8),1}^2= (Z^8_{\sigma(\varphi),1})^{-1}
	\left({1\over 2}(T^{S(P)})^{-1}_{88} - 8I_1(m_u) \right),
     \nonumber \\
     && M_{\sigma^9(\varphi^9),1}^2= (Z^9_{\sigma(\varphi),1})^{-1}
	\left({1\over 2}(T^{S(P)})^{-1}_{99}-8I_1(m_s) \right), \nonumber\\
     && M_{\sigma^8(\varphi^8),2}^2=(Z^8_{\sigma(\varphi),2})^{-1}
	\left({1\over 2G} - 8I_1^{ff}(m_u) \right),\\
     && M_{\sigma^9(\varphi^9),2}^2=(Z^9_{\sigma(\varphi),2})^{-1}
	\left({1\over 2G}-8I_1^{ff}(m_s) \right). \nonumber
     \label{Mpuu}
\ea
In the case of isoscalar mesons it is convenient to combine the scalar and pseudoscalar
fields into 4-vectors
\be
     \Phi=(\varphi_{1}^{8\,r},\varphi_{2}^{8\,r},
	\varphi_{1}^{9\,r},\varphi_{2}^{9\,r}), \qquad
     \Sigma=(\sigma_{1}^{8\,r},\sigma_{2}^{8\,r},
	\sigma_{1}^{9\,r},\sigma_{2}^{9\,r}),
\ee
and introduce  $4\times 4$ matrix functions ${\cal K}_{\sigma(\varphi),ij}$,
instead of old $K_{\sigma(\varphi),ij}^{[a,b]}$, 
where indices $i,j$ run from 1 through 4. This allows us to
rewrite the free part of the effective Lagrangian, which then, with 
the meson fields renormalized, looks 
as follows
\ba
     {\cal L}^{(2)}(\Sigma,\Phi)=\frac12\sum_{i,j=1}^4
     \left(\Sigma_i {\cal K}_{\sigma,ij}(P)\Sigma_j+
     \Phi_i {\cal K}_{\varphi,ij}(P)\Phi_j\right).
	\label{newL2}
\ea
and the functions  ${\cal K}_{\sigma(\varphi),ij}$ are 
\ba
     {\cal K}_{\sigma(\varphi),11}(P)&=&P^2-(m_u\pm m_u)^2-M_{\sigma^8(\varphi^8),1}^2,\nonumber\\
     {\cal K}_{\sigma(\varphi),22}(P)&=&P^2-(m_u\pm m_u)^2-M_{\sigma^8(\varphi^8),2}^2,\nonumber\\
     {\cal K}_{\sigma(\varphi),33}(P)&=&P^2-(m_s\pm m_s)-M_{\sigma^9(\varphi^9),1}^2,\nonumber\\
     {\cal K}_{\sigma(\varphi),44}(P)&=&P^2-(m_s\pm m_s)^2-M_{\sigma^9(\varphi^9),2}^2,\\
     {\cal K}_{\sigma(\varphi),12}(P)&=&{\cal K}_{\sigma(\varphi),21}(P)=\Gamma_{\sigma_u(\eta_u)}(P^2-(m_u\pm m_u)),\nonumber\\
     {\cal K}_{\sigma(\varphi),34}(P)&=&{\cal K}_{\sigma(\varphi),43}(P)=\Gamma_{\sigma_s(\eta_s)}(P^2-(m_s\pm m_s)),\nonumber\\
     {\cal K}_{\sigma(\varphi),13}(P)&=&{\cal K}_{\sigma(\varphi),31}(P)=
	(Z^8_{\sigma(\varphi),1}Z^9_{\sigma(\varphi),2})^{-1/2}(T^{S(P)})^{-1}_{89}. \nonumber
\ea
Now,  to transform (\ref{newL2}) to conventional form, one
should just diagonalize a 4-dimensional matrix, which is better to
do numerically.

\section{The calculation of the amplitudes for the decays of the 
excited scalar meson $\hat a_0$}
Here we collect some instructive formulae, which display a part of
details of the calculations made in this work. Let us demonstrate
how the amplitude of the decay $\hat a_0\to\eta\pi$ is obtained.
The mixing coefficients are taken from  Table~\ref{mixingTable}.
Moreover, the diagrams where  pion vertices contain form factors are
neglected because, as one can see from Table~\ref{mixingTable},
their contribution is significantly reduced.
\ba
      T^{(1)}_{\hat a_0\to\eta\pi}&=&4 \frac{m_u^2}{F_\pi}\biggl\{
	0.82\cdot0.71\cdot Z^{-1/2}\frac{I_2(m_u)}{I_2(m_u)}-\nonumber\\
	&&\left(1.17\cdot 0.71\cdot Z^{-1/2}-0.82\cdot 0.11\right)
	\frac{I_2^f(m_u)}{\sqrt{I_2(m_u)I_2^{ff}(m_u)}}-\nonumber\\
	&&1.17\cdot0.11\cdot\frac{I_2^{ff}(m_u)}{I_2^{ff}(m_u)}\biggr\}\approx0.2\gev,
\ea
\ba
     T^{(2)}_{\hat a_0\to\eta\pi}&=&
     2\frac{m_u^2}{F_\pi}(M_{a_0}^2-M_{\eta}^2-M_{\pi}^2)
     \biggl\{
        0.82\cdot0.71 Z^{-1/2}\frac{I_3(m_u)}{I_2(m_u)}-\nonumber\\
	&&\left(1.17\cdot 0.71\cdot Z^{-1/2}-0.82\cdot 0.11\right)
	\frac{I_3^f(m_u)}{\sqrt{I_2(m_u)I_2^{ff}(m)}}-\nonumber\\
	&&1.17\cdot0.11\frac{I_3^{ff}(m_u)}{I_2(m_u)}\biggr\}\approx3.5 \gev.
\ea
The decay width  thereby is
\be
     \Gamma_{\hat a_0\to\eta\pi}=
     \frac{|T_{\hat a_0\to\eta\pi}|^2}{16\pi M_{\hat a_0}^3}
     \sqrt{M_{\hat a_0}^4\!+\!M_{\eta}^4\!+\!M_{\pi}^4\!-\!
     2(M_{\hat a_0}^2 M_{\eta}^2\!+\!M_{\hat a_0}^2 M_{\pi}^2\!+\!M_{\eta}^2 M_{\pi}^2)}\approx
     160 \mev.
\ee
Here $I_2(m_u)=0.04$, $I_2^f(m_u)=0.014c$, $I_2^{ff}(m_u)=0.015c^2$,
$I_3(m_u)=0.11 \gev^{-2}$, $I_3^f(m_u)=0.07c\gev^{-2}$,$I_3^{ff}(m_u)=0.06c^2\gev^{-2}$
and $c$ is the external form factor parameter factored out
and cancelled in the ratios of the integrals.

For the decay into strange mesons we obtain (see Fig.1)
\ba
     &&\!\!\!\!T_{\hat a_0\to K^+K^-}\!=\!
     C_K\!\left(-\frac{iN_c}{16\pi^2}\right)\!\!\int\!d^4k
	{ \Tr[(m_u+\Slash{k}+\Slash{p}_1)\gamma_5(m_s+\Slash{k})\gamma_5(m_u+\Slash{k}-\Slash{p}_2)]
      \over
	(m_s^2-k^2)(m_u^2-(\Slash{k}-\Slash{p}_1)^2)(m_u^2-(\Slash{k}-\Slash{p}_2)^2)
	}\approx\nonumber\\
     &&\!\!\!\!2C_K\left\{
	 (m_s+m_u)I_2(m_u)-\Delta I_2(m_u,m_s)-
	 [m_s(M_{\hat a_0}^2-2M_{K}^2)-\right.\\
     &&\!\!\!\!\left.\quad	2\Delta^3]I_3(m_u,m_s)\nonumber
        \right\},
\ea
where $\Delta=m_s-m_u$ and
\be
	I_3(m_u,m_s)=-i\frac{N_c}{(2\pi)^4}\int_{\Lambda_3}\!\frac{d^4k}{(m_u^2-k^2)^2(m_s^2-k^2)}.
\ee
The coefficient $C_K$ absorbs the Yukawa coupling constants and some structure coefficients.
The integral $I_2(m_u,m_s)$ is defined by (\ref{DefIf}).
This is only the part of the amplitude without form factors.
The complete amplitude of this process is a sum of contributions which
contain also the integrals $I_2^{f..f}$ and $I_3^{f..f}$ with form factors. 
Thus, the amplitude is
\ba
	&&T_{\hat a_0\to K^+K^-}=T^{(1)}+T^{(2)},\\
	&&T^{(1)}=\frac{m_u+m_s}{2F_K}
	\bigl\{
	(m_s+m_u)\cdot 0.13-\Delta\cdot 0.21\bigr\}\approx 0.2\gev,\\
	&&
	T^{(2)}=\frac{m_u+m_s}{2F_K}\bigl\{[m_s(M_{a_0}^2-2M_{K}^2)-2\Delta^3]\cdot1\gev^{-2}\bigr\}\approx 2.3\gev
	,\\
	&& F_K=1.2F_\pi.\nonumber
\ea
The decay width therefore is evaluated to be
\be
	\Gamma_{\hat a_0\to K^+K^-}=\Gamma_{\hat a_0\to \bar K^0K^0}\approx 50 \mev.
\ee



\newpage
\begin{figure}[h]
\begin{center}
\caption{  Diagrams describing decays of $\hat a_0$ to pseudoscalars.}
\psfig{file=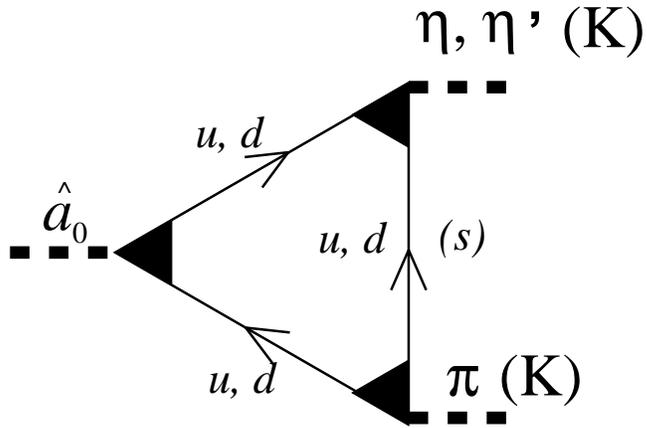}
\end{center}
\end{figure}

\end{document}